\documentclass[twocolumn,aps,showpacs]{revtex4-1}
\usepackage[latin9]{inputenc}
\usepackage{amsmath}
\usepackage{amssymb}
\usepackage{graphicx}

\makeatletter

\newcommand*\LyXThinSpace{\,\hspace{0pt}}
\DeclareTextSymbolDefault{\textquotedbl}{T1}

\usepackage{color}

\def\NOT(#1,#2){\OneQubitGate(#1,#2){$X$}}

\makeatother

\begin{document}

\title{Fast Quantum State Tomography in the Nitrogen Vacancy Center of Diamond}

\author{Jingfu Zhang, Swathi S. Hegde and Dieter Suter\\
 Fakultaet Physik, Technische Universitaet Dortmund,\\
 D-44221 Dortmund, Germany}

\date{\today}
\begin{abstract}
Quantum state tomography (QST) is the procedure for reconstructing
unknown quantum states from a series of measurements of different
observables. Depending on the physical system, different sets of observables
have been used for this procedure. In the case of spin-qubits, the
most common procedure is to measure the transverse magnetization of
the system as a function of time. Here, we present a different scheme
that relies on time-independent observables and therefore does not
require measurements at different evolution times, thereby greatly
reducing the overall measurement time. To recover the full density
matrix, we use a set of unitary operations that transform the density
operator elements into the directly measurable observable. We demonstrate
the performance of this scheme in the electron-nuclear spin system
of the nitrogen vacancy center in diamond. 
\end{abstract}

\pacs{03.67.Pp,03.67.Lx}

\maketitle
\textit{Introduction}.\textendash Finding the state of a quantum system
is one of the main tasks for many applications in basic quantum physics
\cite{quantumOptics} as well as in many emerging quantum technologies,
such as in the field of quantum information \cite{nielsen,Stolze:2008xy}.
Using quantum state tomography (QST), one can reconstruct the quantum
state represented by a density operator, which contains the full information
about the system \cite{nielsen,LecQST,ALTEPETER2005105}. Performing
a full QST requires serial measurements of a complete set of observables.
The size of such a complete set and therefore the number of individual
measurements and the measurement time all grow exponentially with
the number of qubits in the system.

While QST has been performed for many years \cite{1019,1103}, it
was often done in the form of time-dependent measurements \cite{PhysRevA.69.052302,6991}.
In spin-based systems, such as nuclear magnetic resonance \cite{Ernst,Stolze:2008xy},
it is often not possible or not optimal to use projective measurements.
Instead, the established procedure relies on the measurement of free
induction decays (FIDs) \cite{1019,1103}, which may generate information
on multiple density operator elements in a single scan \cite{PhysRevA.69.052302,RevModPhys.76.1037,Vandersypen:2007fk}.
This procedure has therefore been well established in ensemble quantum
information processing.

In the case of single spin qubits, such as the nitrogen vacancy (NV)
center in diamond \cite{Doherty:2013uq,Suter201750}, this type of
measurement is also applicable \cite{6991,sci08}, but the precessing
transverse magnetization that is detected in a conventional FID experiment,
is not directly observable. Instead, the transverse magnetization
(coherence) has to be converted into population of the electron spin
and detected as a change of the photoluminescence count rate; this
is known as the Ramsey-fringe method \cite{Suter201750}. Since this
type of readout results in a large number of individual measurements,
the procedure becomes even more time-consuming. Therefore, we propose
here a different approach that is significantly more efficient for
this type of qubit: eliminating the need for free evolution reduces
the number of actual measurements by several orders of magnitude,
with a corresponding reduction of the overall measurement time.

In our scheme, the photon count rate is the immediate observable;
it is directly connected to populations of the electron spin, which
correspond to the diagonal elements or their linear combination in
the density matrix. Since such populations do not change during free
evolution, the observable is not time dependent. The other elements
of the density matrix can be transformed to the observable through
unitary transformations. With a suitable decomposition of the density
operator, every element in the basis set can be converted into the
observable and therefore be read out by a single measurement. Overall,
this procedure provides a dramatic speed-up by several orders of magnitudes,
compared to the measurement of precessing magnetization. As an experimental
demonstration, we implement this procedure in the single NV center
of diamond, which is used in many emerging applications of quantum
information and sensing technologies. \cite{ladd2010quantum,blencowe2010quantum,cai2014hybrid,kurizki2015quantum,Suter201750,RevModPhys.92.015004}.

\textit{Single-qubit state tomography}.\textendash We start with the
QST of single qubit \cite{Howard_2006,PhysRevLett.112.050502,PhysRevA.69.032307,PhysRevA.77.022307}.
We consider systems where the measurement of diagonal density operator
elements is easy, such as in the NV centers of diamond where the populations
can be determined by photon counting \cite{Suter201750}. In the case
of single qubits, the relevant Hilbert space is spanned by the computational
basis $\{|0\rangle,|1\rangle\}$, which are the eigenstates of the
Pauli operator $Z$ with eigenvalues $\pm1$. The density matrix describing
the quantum state can be expanded in terms of the unit operator $E$
and the Pauli matrices $(X,Y$ and $Z)$ as 
\begin{equation}
\rho=c_{E}E+c_{X}X+c_{Y}Y+c_{Z}Z,\label{eq:E}
\end{equation}
where $c_{E}=1/2$ for a normalized density operator and the other
$c_{i}$ are the weights of the corresponding Pauli matrices. The
diagonal elements $\rho_{11}$ and $\rho_{22}$, which correspond
to the populations $p_{|0\rangle}$ and $p_{|1\rangle}$ of the states
$|0\rangle$ and $|1\rangle$, are related to the coefficients $c_{E}$
and $c_{Z}$ as 
\begin{equation}
c_{E}=1/2=(p_{|0\rangle}+p_{|1\rangle})/2,\quad c_{Z}=(p_{|0\rangle}-p_{|1\rangle})/2.\label{eq:nonideal-1}
\end{equation}
To measure the off-diagonal elements of the density operator $\rho$,
we apply operations $X_{90}$ and $Y_{90}$ to transform them to diagonal
elements. Here $X_{\alpha}$ and $Y_{\alpha}$ are rotations of the
qubit around the $x-$ and $y-$ axis by an angle $\alpha$. They
transform $c_{Y}Y$ and $c_{X}X$ to $c_{Y}Z$ and $-c_{X}Z$, respectively.
Therefore $c_{X}$ and $c_{Y}$ can be measured directly in the transformed
states. 
\begin{figure}
\begin{centering}
\includegraphics[width=7cm]{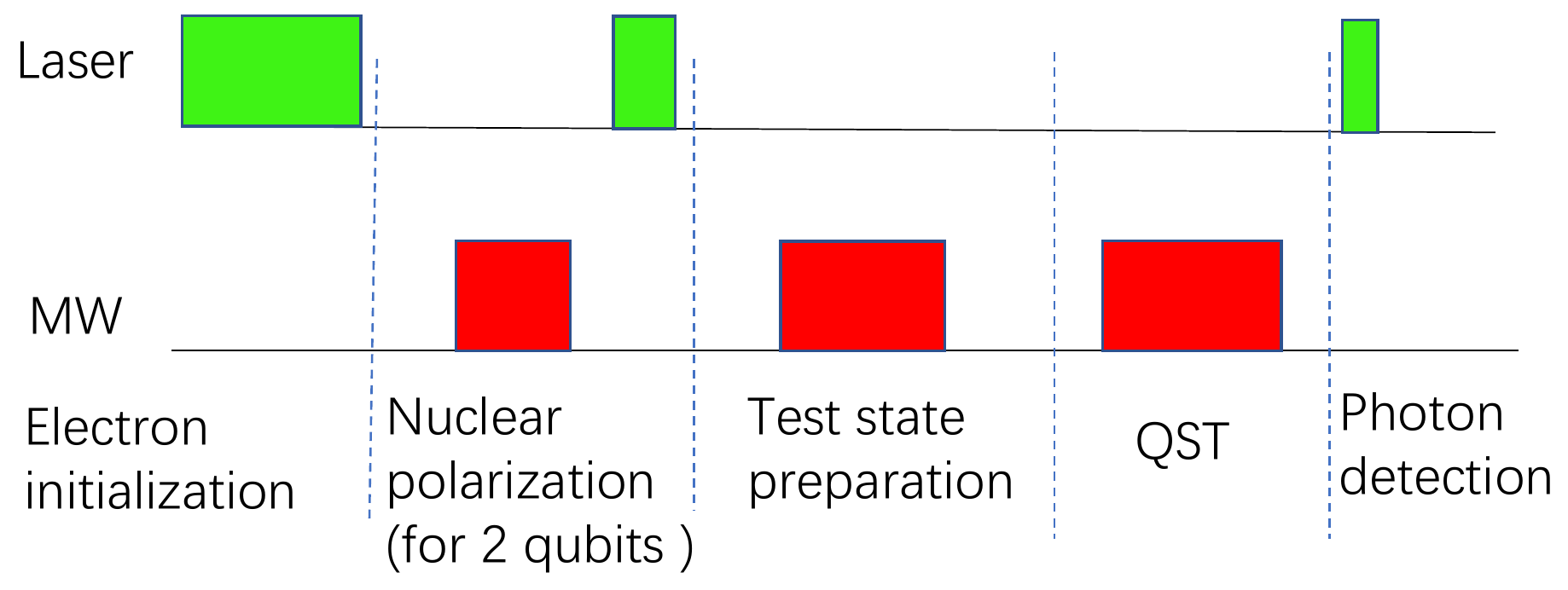} 
\par\end{centering}
\centering{}\caption{Pulse sequence for state preparation followed by QST. The nuclear
spin polarization step is only used for the 2 qubit system. Each green
box represents a laser pulse, while the red boxes represent sequences
microwave (MW) pulses. \label{pulse}}
\end{figure}

For the experimental demonstration, we used the electron spin of a
single NV center in a diamond sample with natural abundance ($\sim1.1$\%)
of $^{13}$C. The experiments were performed at room temperature.
The static magnetic field $B$ was aligned along the symmetry axis
of the NV center. The relevant Hamiltonian of the electron spin is
then $\mathcal{H}_{e}/(2\pi)=DS_{z}^{2}-B\gamma_{e}S_{z}$. Here $S_{z}$
denotes the spin-1 operator for the electron, $D$ the zero-field
splitting and $\gamma_{e}$ the gyromagnetic ratio \cite{Suter201750}.
Fig. \ref{pulse} shows the pulse sequence, which always starts with
the polarization of the electron spin into the state $m_{S}=0$, using
a pulse of a 532 nm laser. The polarization is higher than $70\%$
\cite{Howard_2006}, and in the present work we can approximate the
pseudopure state as a pure state $|0\rangle$ as discussed in the Supplementary Material 
(SM, section IC) \cite{SMQST}.

The population of the state $m_{S}=0$ can be measured by the count
rate $r$ of the fluorescence detected during a second laser pulse,
since the state $m_{S}=0$ fluoresces more strongly than $m_{S}=\pm1$
\cite{doi:10.1002/pssa.200671403,Childress281,Suter201750,Howard_2006}:
\begin{equation}
r=r_{min}+p_{|0\rangle}(r_{max}-r_{min}).\label{eq:cr_general}
\end{equation}
The maximum count rate $r_{max}$ corresponds to the system being
in state $m_{S}=0$, while the minimum count rate $r_{min}$ results
when the system is in $m_{S}=\pm1$. The readout is destructive, since
the laser pumps the system back to the state $|0\rangle$. Therefore,
the measurement time has to be kept relatively short \cite{Suter201750}
to obtain a good measure of the instantaneous population. To reduce
the effects of drift and laser power fluctuations, we always calibrate
the count rate against a measurement of $r_{max}$ obtained after
re-pumping the system to the $m_{S}=0$ state.

The single qubit is obtained from the electron spin states $m_{S}=0$
and $m_{S}=-1$, which we identify with the two logic states $|0\rangle$
and $|1\rangle$. For the 1-qubit case, the test state preparation
and the transformations for the QST was implemented by  single microwave
(MW) pulses with a Rabi frequency of 9 MHz. The measured count rate
$r$ then allows us to determine the populations of the state as 
\begin{equation}
p_{|0\rangle}=(r-r_{min})/\delta_{r},\quad p_{|1\rangle}=(r_{max}-r)/\delta_{r},\label{eq:idealp01}
\end{equation}
where $\delta_{r}\equiv r_{max}-r_{min}$. Writing $r_{N}$, $r_{X}$,
and $r_{Y}$ for the count rates measured after the 3 operations NOOP
(no operation), $X_{90}$ and $Y_{90}$ we obtain the coefficients
\begin{eqnarray}
c_{X} & = & (1/2)-(r_{Y}-r_{min})/\delta_{r}\nonumber \\
c_{Y} & = & (r_{X}-r_{min})/\delta_{r}-(1/2)\nonumber \\
c_{Z} & = & (r_{N}-r_{min})/\delta_{r}-(1/2).\label{eq:cz-1-1-1}
\end{eqnarray}

To test the QST procedure, we first prepared test states $|0\rangle$,
$|1\rangle$, $|+\rangle=(|0\rangle+|1\rangle)/\sqrt{2}$, and $|-\rangle=(|0\rangle-i|1\rangle)/\sqrt{2}$
by applying the operations NOOP, $X_{180}$, $Y_{90}$ and $X_{90}$
to state $|0\rangle$, respectively. These states are eigenstates
of the Pauli matrices $Z$, $X$, and $Y$.

\begin{figure}
\begin{centering}
\includegraphics[width=7cm]{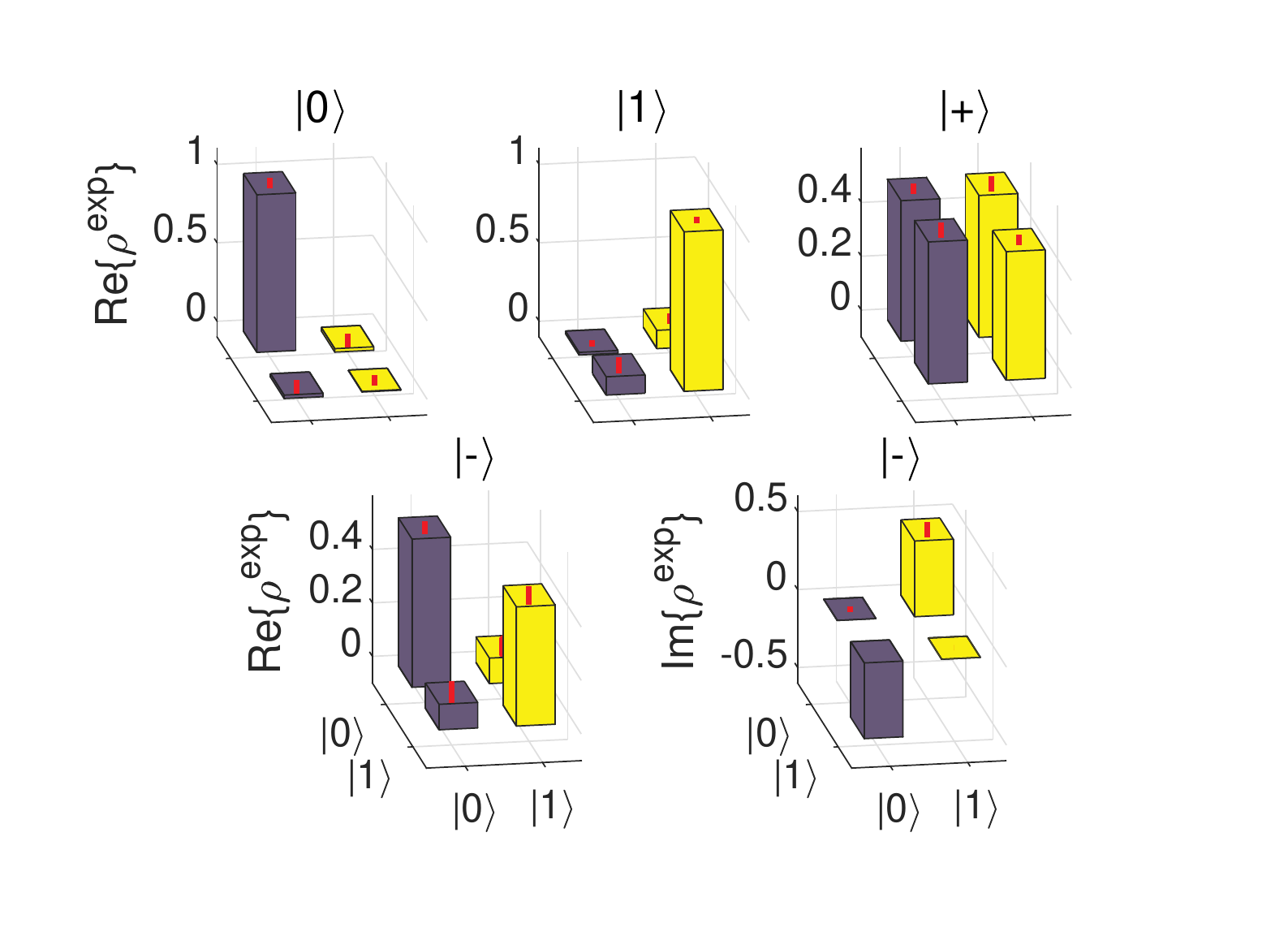} 
\par\end{centering}
\centering{}\caption{Experimental results of the tomography of a single qubit. The top
row shows the real parts of the measured density matrices for states
$|0\rangle$, $|1\rangle$ and $|+\rangle$, as indicated in the panel.
The bottom row shows the real and imaginary parts for state $|-\rangle$.
The error bars show the standard deviations obtained by repeating
the measurements. \label{tomosingle}}
\end{figure}

Fig. \ref{tomosingle} shows a graphical representation of the density
matrices, reconstructed from the measured coefficients, whose values
are given in the  SM \cite{SMQST}. The top row shows the
real parts of the density matrices for states $|0\rangle$, $|1\rangle$
and $|+\rangle$. The bottom row shows the real and imaginary parts
for state $|-\rangle$.

From the reconstructed density operators, we calculated the fidelity
\cite{Wang08} 
\begin{equation}
F=\frac{Tr\{\rho^{th}\rho^{exp}\}}{\sqrt{Tr\{\rho^{th}\rho^{th}\}Tr\{\rho^{exp}\rho^{exp}\}}},\label{eq:fid}
\end{equation}
where $\rho^{th}$ and $\rho^{exp}$ are the theoretical and experimentally
reconstructed density operators. The resulting fidelities for the
input states $|0\rangle$, $|1\rangle$, $|+\rangle$ and $|-\rangle$
are 0.994, 0.985, 0.995 and 0.986, respectively. The deviation between
experiment and theory can be mainly attributed to the statistics of
the photon detection and the control errors (pulse imperfection) in
implementing the state preparation and tomography. Compared with the
fidelity for state $|0\rangle$, the slightly lower fidelities for
the other three states can be used to estimate the process fidelity
of the state preparation.

For a second example of 1-qubit QST, we identified the qubit again
with the same electronic spin states $m_{S}=0$ and $m_{S}=-1$, but
now conditional on the $^{14}$N nuclear spin being in the $m_{N}=1$
state. The experiments were performed in a center in the $^{12}$C
enriched ($99.995\%$) diamond sample \cite{1882-0786-6-5-055601,PhysRevLett.110.240501,doi:10.1063/1.4731778}.
We obtained a slightly higher fidelity than the earlier experiments.
The results are presented in the SM \cite{SMQST}.

\textit{State tomography for 2 qubits}.\textendash Moving to a 2-qubit
system, we use the basis states $\{|00\rangle,|01\rangle,|10\rangle,|11\rangle\}$
and we expand the density operator in a basis of products of single
qubit operators: 
\begin{equation}
\rho=\sum_{m,n=1}^{4}c_{mn}a_{m}\otimes b_{n}\label{den2}
\end{equation}
where $a_{m},b_{n}\in\{E,X,Y,Z\}$ represent the unit operator $E$
and the Pauli matrices acting on one of the qubits. Since the trace
of a normalised density operator is unity, the coefficient $c_{EE}=1/4$
is fixed. The goal of the tomography is to determine the other 15
coefficients $c_{mn}$.

The primary observable for measuring populations is again the photon
count rate, which depends on the state of the electron spin before
the readout pulse is applied, but is independent of the state of the
nuclear spin. The measured count rate is then, in analogy to Eq. \eqref{eq:cr_general}
\begin{equation}
r=r_{min}+(p_{|00\rangle}+p_{|01\rangle})(r_{max}-r_{min}).\label{eq:cr_gerenal2id}
\end{equation}
Eq. \eqref{eq:idealp01} also holds, with $p_{|0\rangle}\to p_{|00\rangle}+p_{|01\rangle}$
and $p_{|1\rangle}=p_{|10\rangle}+p_{|11\rangle}$. Therefore

\begin{eqnarray}
p_{|00\rangle}+p_{|01\rangle} & =(r_{N}-r_{min})/\delta_{r}\label{eq:cr_2Nid}\\
p_{|10\rangle}+p_{|11\rangle} & =(r_{max}-r_{N})/\delta_{r}\label{eq:cr_2_180id}
\end{eqnarray}
and 
\begin{equation}
c_{ZE}=(r_{N}-r_{min})/(2\delta_{r})-(1/4),\label{eq:diagTRid-1}
\end{equation}
which corresponds to $c_{Z}$ in Eq. (\ref{eq:cz-1-1-1}) in the single
qubit QST.

To determine the remaining coefficients of the density operator, we
apply a set of unitary operations $R$ to transform the relevant operators
$c_{mn}a_{m}b_{n}$ to $c_{mn}ZE$. The coefficient $c_{mn}$ in the
transformed density matrix can be directly measured using Eq. (\ref{eq:diagTRid-1}),
by replacing $r_{N}$ by $r_{R}$, where $r_{R}=2\delta_{r}[c_{mn}+(1/4)]+r_{min}$
denotes the count rate measured from the transformed density matrix.
Overall we can use 15 measurements to obtain the 15 coefficients,
i.e., one measurement for each element of the density operator.

To demonstrate the 2-qubit scheme, we used the electron spin coupled
to a single $^{13}$C nuclear spin, where the electron spin in states
$m_{S}=0$ and $m_{S}=-1$ was assigned as qubit 1 and $^{13}$C nuclear
spin qubit 2. We used a $^{12}$C enriched ($99.995\%$) diamond to
minimize decoherence due to additional $^{13}$C nuclear spins \cite{doi:10.1063/1.4731778}.
In this context, we focus on the electron and $^{13}$C subsystem
with the $^{14}$N in the state $m_{N}=+1$. The pulse sequence is
shown in Fig. \ref{pulse}, with more details given in the SM \cite{SMQST}. The
required operations can be efficiently generated by applying a small
number of MW pulses acting on the electron spin, combined with free
precession \cite{PhysRevA.76.032326,PhysRevA.78.010303,zhang18,swathi19,zhang19,PhysRevLett.128.230502}.

To prepare the pure state $|00\rangle$, we first polarized the electron
spin, swapped the states of the two qubits and re-polarized the electron
spin \cite{zhang18,swathi19}. Additional details are provided in
the SM \cite{SMQST}. We implemented $X_{90}\otimes E$, $Y_{90}\otimes E$ and
$X_{180}\otimes E$ by single MW pulses. The other required unitaries
were implemented by pulse sequences that were designed by optimal
control (OC) theory \cite{Mitchell:1998:IGA:522098,zhang18}. These
pulse sequences transfer the target operators to $ZE$ with fidelities
of $\geq0.99$. The pulse sequences consist of up to 3 MW pulses and
the same number of free precession periods and total durations up
to 15 $\mu$s, which is short compared to the transverse relaxation
times $T_{2}=700$ $\mu$s and $T_{2}^{*}=40$ $\mu$s of the electron
spin. Additional details are given in the SM \cite{SMQST}.

As experimental demonstrations, we reconstruct the density matrices
of the following states: $s_{1}=|00\rangle$, $s_{2}=|0\rangle(|0\rangle+|1\rangle)/\sqrt{2}$,
$s_{3}=(|00\rangle+|11\rangle)/\sqrt{2}$ and $s_{4}=(|01\rangle+|10\rangle)/\sqrt{2}$.
The states $s_{2}$- $s_{4}$ were generated by applying sequences
of MW pulses and delays to $|00\rangle$. Each sequence consists of
3 pulses and 3 delays. The theoretical fidelity of the generated state
is $>0.99$.

\begin{figure}
\centering{}\includegraphics[width=1\columnwidth]{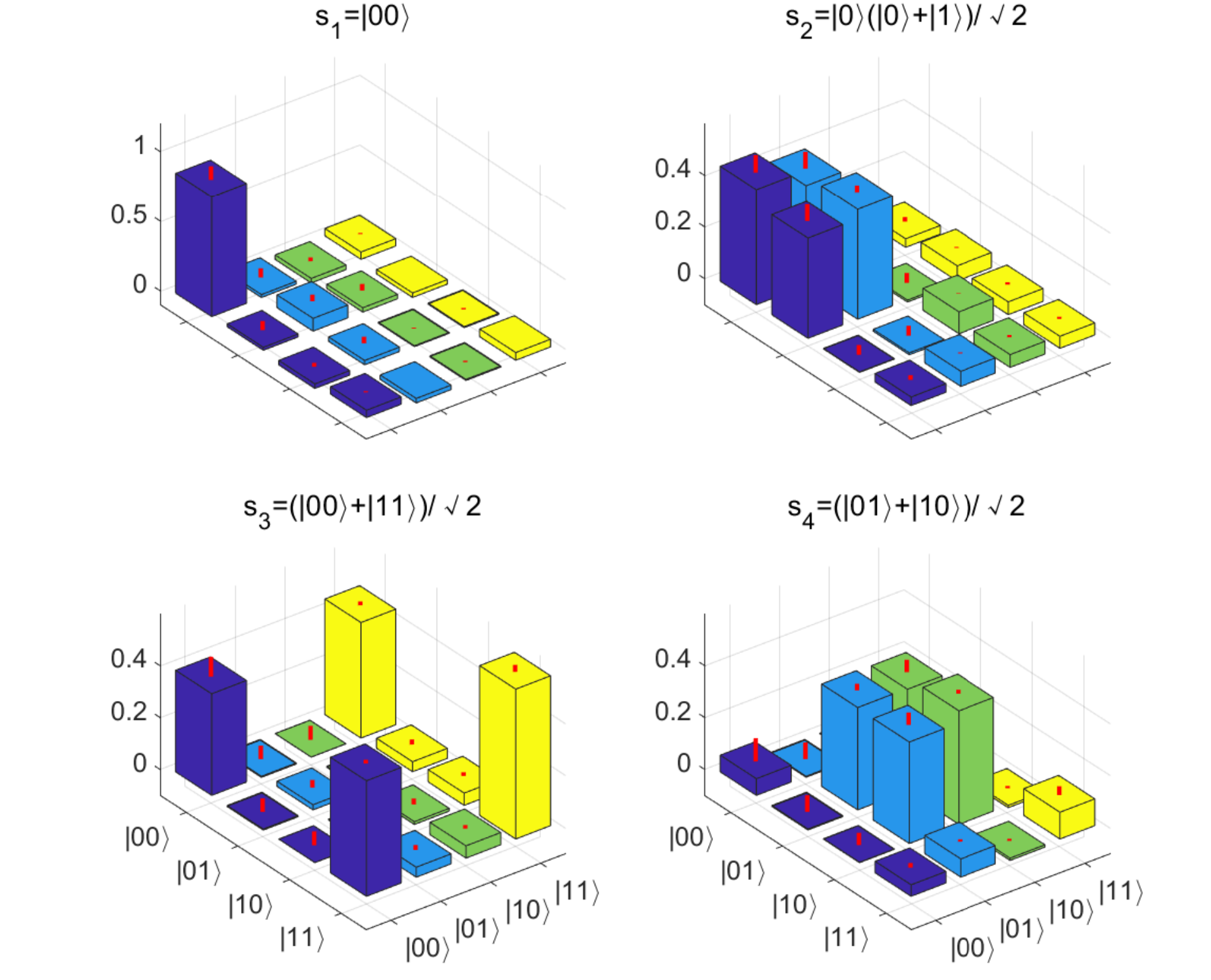}
\caption{Real parts of the density matrices experimentally reconstructed by
the QST for the input states $s_{1}$- $s_{4}$. \label{real}}
\end{figure}

The experimental results for the real parts of the measured density
matrices are illustrated in Fig. \ref{real}. The root-mean-square
(RMS) values of imaginary parts in the experimental density operator
are 0.028, 0.033, 0.039 and 0.026, for the input states $s_{1}$-
$s_{4}$, respectively. We present the measured imaginary parts of
the density matrices in the SM \cite{SMQST}. 

The experimental fidelities for the states $s_{1}$ - $s_{4}$, are
0.98, 0.97, 0.97, and 0.97, respectively. The main contributions to
the deviation from unity are i) dephasing (0.05\%), ii) the theoretical
imperfections of the pulse sequences (1\%), iii) experimental imperfections
of the MW pulses (1\%) and iv) photon counting statistics (2\%). Additional
details are presented in the SM (Section VC) \cite{SMQST}. We are currently optimizing
the conversion sequences such that they combine high fidelity for
the unitary conversion operation with suppression of dephasing \cite{PhysRevA.76.032326,Pan13}.

\textit{Discussion.\textendash }Our scheme can be straightforwardly
generalized to the multiple qubit system. In a $n$ qubit system,
the observable is $Z\underbrace{E...E}_{n-1}$, denoting a product
operator with $Z$ for the electron spin qubit and $E$ for the $n-1$
nuclear spin qubits. Eq. (11) is generalized to 
\begin{equation}
c_{Z\underbrace{E...E}_{n-1}}=(r_{N}-r_{min})/(2^{n-1}\delta_{r})-(1/2^{n}).
\end{equation}
In a similar way, all product operators can be transformed to $Z\underbrace{E...E}_{n-1}$
by unitary operations. Therefore we need $(2^{2n}-1)$ measurements
for reconstructing the full density operator. More details are presented
in the SM (Section VE) \cite{SMQST}. The number of measurements required by the
time-dependent experiments (Ramsey) also increases proportional to
the number of elements in the density operator. While the precise
number depends on the details of the coupling network, additional
couplings allow one to extract more density matrix elements from a
single FID measurement and therefore reduce the number of FIDs that
must be measured \cite{PhysRevA.69.052302}. On the other hand, they
lead to increased spectral crowding, which requires a larger number
of points per FID. As a result, the time saving of the time-independent
over the time-dependent approach does not depend on the number of
qubits. More details are shown in the SM (Section VE) \cite{SMQST}. We therefore
conclude that the time saving of \textgreater{} 2 orders of magnitude
should be similar for all relevant quantum registers. However, full
QST for systems with \textgreater 3 qubits will probably remain impractical
even with this faster method.

\textit{Conclusion}.\textendash Quantum state tomography is an essential
tool for the analysis of quantum mechanical systems as it allows one
to extract all available information \cite{quantumOptics,nielsen,Stolze:2008xy}.
Accordingly, efficient procedures for QST are valuable for a vast
range of applications where information on multiple density operator
elements is accessed \cite{note}. Early QST experiments, e.g. in
quantum optics \cite{LecQST,ALTEPETER2005105} were based on measurements
in different bases to extract the coefficients of the density operator
components. In the system that we are considering, only a single observable
is available. To access different density operator components, we
therefore convert them into the available observable through a set
of unitary transformations. Early QST experiments by liquid-state
NMR \cite{LNMR}, also used a single measurement basis but since the
relevant measurement is not projective, it was possible to continuously
monitor the time-evolution of the density operator, which converts
density operator components that are not directly observable into
the observable one \cite{1019,LNMR}. In the case of QST of single
spins in solids, the evolution of coherences can not be observed directly;
it was therefore replaced by indirect detection using the Ramsey method
\cite{Ramsey50}. While this approach allowed one to transfer the
techniques developed for liquid-state NMR to the single-spin systems,
it generates a huge overhead, since a single measurement is replaced
by a sequence of typically several hundred measurements with different
evolution times. In the method presented here, we remove this overhead,
which allows a speed-up of the QST by several orders of magnitude
compared to the measurement of time-dependent observables, for both
the number of required measurements and the overall measurement time
(see SM, Section VD) \cite{SMQST}. Similar to the existing procedures, the reconstruction
of the density operator can be improved by combing the measurement
results with statistical inference methods that result in a density
matrix that is close to the physical state \cite{SI1,SI2,SI3,SI4,SI5,SI6}.
Since QST is the main prerequisite for quantum process tomography
(QPT) \cite{PhysRevLett.78.390,doi:10.1080/09500349708231894}, our
method is also very helpful for speeding up QPT. For our experimental
demonstration, we used a nitrogen vacancy center in diamond, but the
scheme should be equally applicable to other physical systems, such
as photons, atomic ensembles and trapped
ions \cite{ALTEPETER2005105,PhysRevLett.105.053201,PhysRevLett.97.220407,Keselman_2011}.

\textit{Acknowledgments}.\textendash This project has received funding
from the European Union's Horizon 2020 research and innovation programme
under grant agreement No 828946. The publication reflects the opinion
of the authors; the agency and the commission may not be held responsible
for the information contained in it. We thank Sven Lauhof for the
help in the experiment.


\begin{thebibliography}{10}
\bibitem{note} Other experiments, such as the measurement of transition
frequencies, do not require full QST.
\end{thebibliography}

\end{document}